\documentclass{svproc}
\usepackage{url}

\usepackage[pdftex]{graphicx}
 \DeclareGraphicsExtensions{.pdf}

\begin{document}
\mainmatter              
\title{Jet Identification with Zest}
\titlerunning{Jet Identification with Zest}  
%
\author{Ankita Budhraja \and Ambar Jain
}
\authorrunning{Budhraja and Jain} 
%
%
\institute{Indian Institute of Science Education and Research, Bhopal MP 462066, India\\
\email{ankitab@iiserb.ac.in}
}

\maketitle              

\begin{abstract}
We present a new observable `zest'  and demonstrate 
its potential to differentiate between jets originated by gluons, top quark and vector bosons. Zest has salient properties such as boost invariance, stability against global color flow of partons and inclusion or exclusion of a few soft particles to the jet. For a gluon jet, zest distribution is also insensitive to the jet mass. We show that when zest is used in conjunction with other observables, it can yield high gluon rejection while retaining high signal sample.
\end{abstract}
%

{\bf Intrduction: }The standard model heavy particles, namely, $W$ and $Z$ bosons, Higgs boson and the top quark, dominantly decay to light quarks and gluons which in turn shower into jets of hadrons. Since jets are invariably produced at LHC, heavy SM particles decaying to jets are often faked by jets from light quarks and gluons (collectively parton jets). One might expect that the jet mass may cut out undesired gluon jets largely, however, as it happens to be the case, gluon jet cross-section has a large and long tail that is worsened by the underlying event 
and the pile-up.
A great deal of effort \cite{all-jet-taggers} is underway in jet identification by constructing observables that can reject the parton jets and thus reduce the background with a goal to improve the signal rate to miss-tag rate ratio. The purpose of this work is to improve upon such strategies by studying a new observable {\em zest} similar to transverse-zeal introduced in context of jet quenching studies \cite{zeal-transverse}. 
%

%
%
{\bf Zest of a Jet: }For a jet composed of hadrons, zest \cite{future} is defined as
\begin{eqnarray}
\label{eq:zest}
\zeta = \frac{-1}{\log \big (\sum_{i \in  \rm{Jet}} e^{-P_T/\vert \vec p_{\perp i} \vert}\big )},
\end{eqnarray}
where $P_T = \sum_{i \in \rm{Jet}} \vert \vec p_{\perp i} \vert$ and $\vec p_{\perp i}$ is the transverse momentum of the $i^{\rm{th}}$ particle in the jet with respect to the jet  axis. Here we have assumed that a jet has been constructed using a suitable jet algorithm with quintessential grooming \cite{jet-grooming}. Zest has some interesting properties: (1) $\zeta$ is invariant under boosts along the jet axis, (2) it is mostly sensitive to particles with large transverse momenta and insensitive to the soft particles in the jet, (3) it is largely insensitive to the global color flow of partons. Due to its little sensitivity to the soft particles in the jet, zest distributions are stable against inclusion or exclusion of a few soft particles \cite{future}. We further expect it to be stable against jet reconstruction methods, and jet grooming techniques. Owing to its salient properties, $\zeta$-distribution of gluons are narrow, independent of jet energy and have little overlap with similar distribution of most other particles. $\zeta$ is highly non-linear and infrared unsafe due to which it cannot be calculated in perturbation theory, however it can be computed using Monte Carlo based event generators. Zest provides a perspective into the jet substructure that may not be accessible with infrared safe observables. Zest is hadronization model dependent but we expect zest based discrimination is not\cite{future}.
%

{\bf Results: }All jets are simulated using Pythia 8\cite{pythia} by inserting particles with energy $500 \,{\rm GeV}$ along the $z$-axis and allowing them to decay to hadronic modes only. For gluon jets, one energetic gluon of energy $500 \,{\rm GeV}$ and off-shellness same as the heavy particle mass was inserted along the $z$-axis, while a soft on-shell gluon of energy $1\,{\rm MeV}$ with opposite color was inserted along the negative $z$-axis to shower a color singlet event. The gluon jet was identified by constructing the thrust axis \cite{thrust} and taking all particles in the forward hemisphere. The resulting jets were accepted if their masses were within $5 \,{\rm GeV}$ of the heavy particle. Likewise, for top quarks, a top anti-top pair of opposite color were inserted back to back  for simulation, whereafter jets were constructed by dividing the event into two hemispheres perpendicular to the thrust axis. Zest distributions for gluons and heavy particles are shown in Fig. 1(a).
\begin{figure}[t]
\centering
\includegraphics[width=0.43\textwidth]{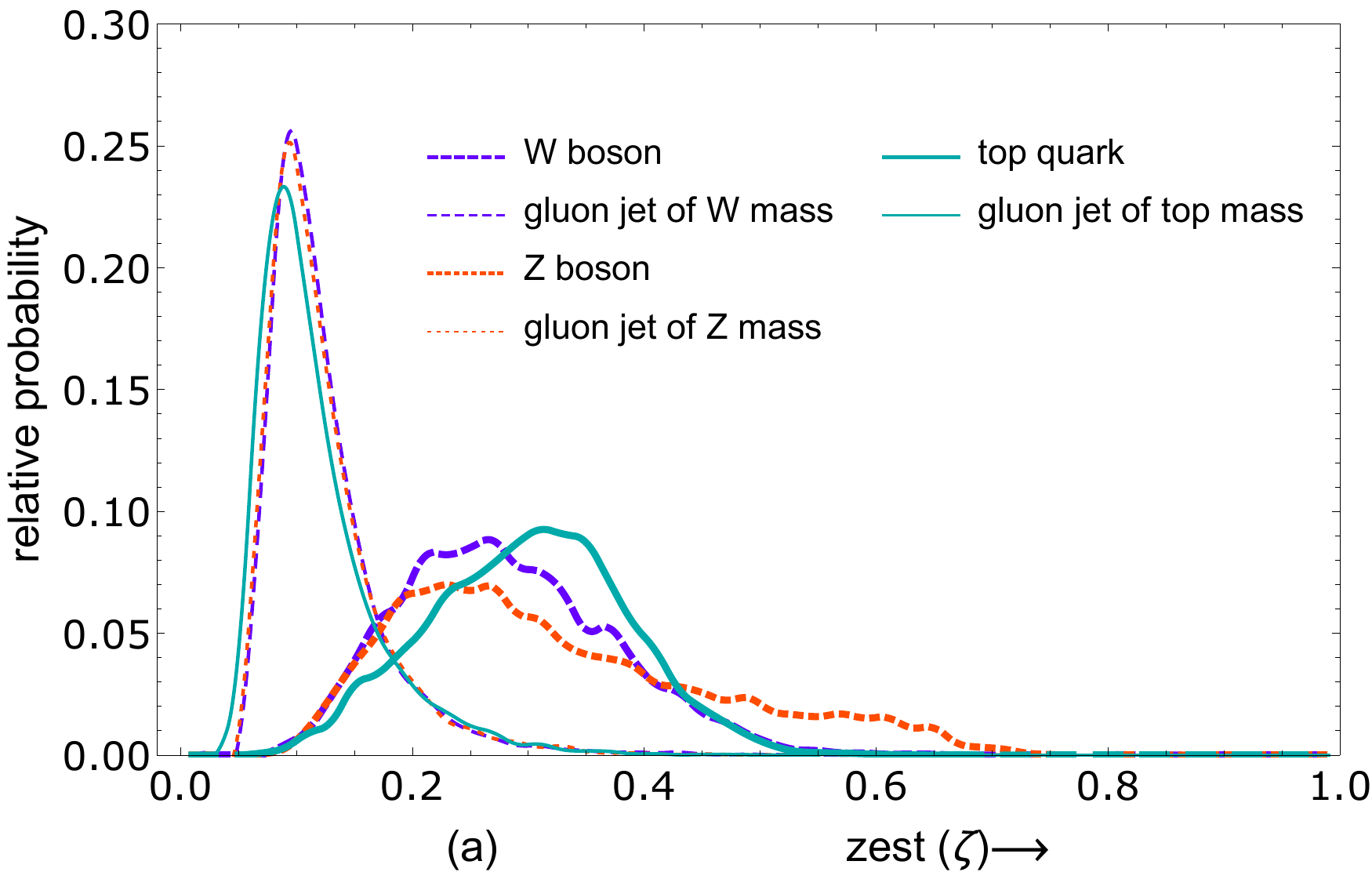} 
\includegraphics[width=0.5\textwidth]{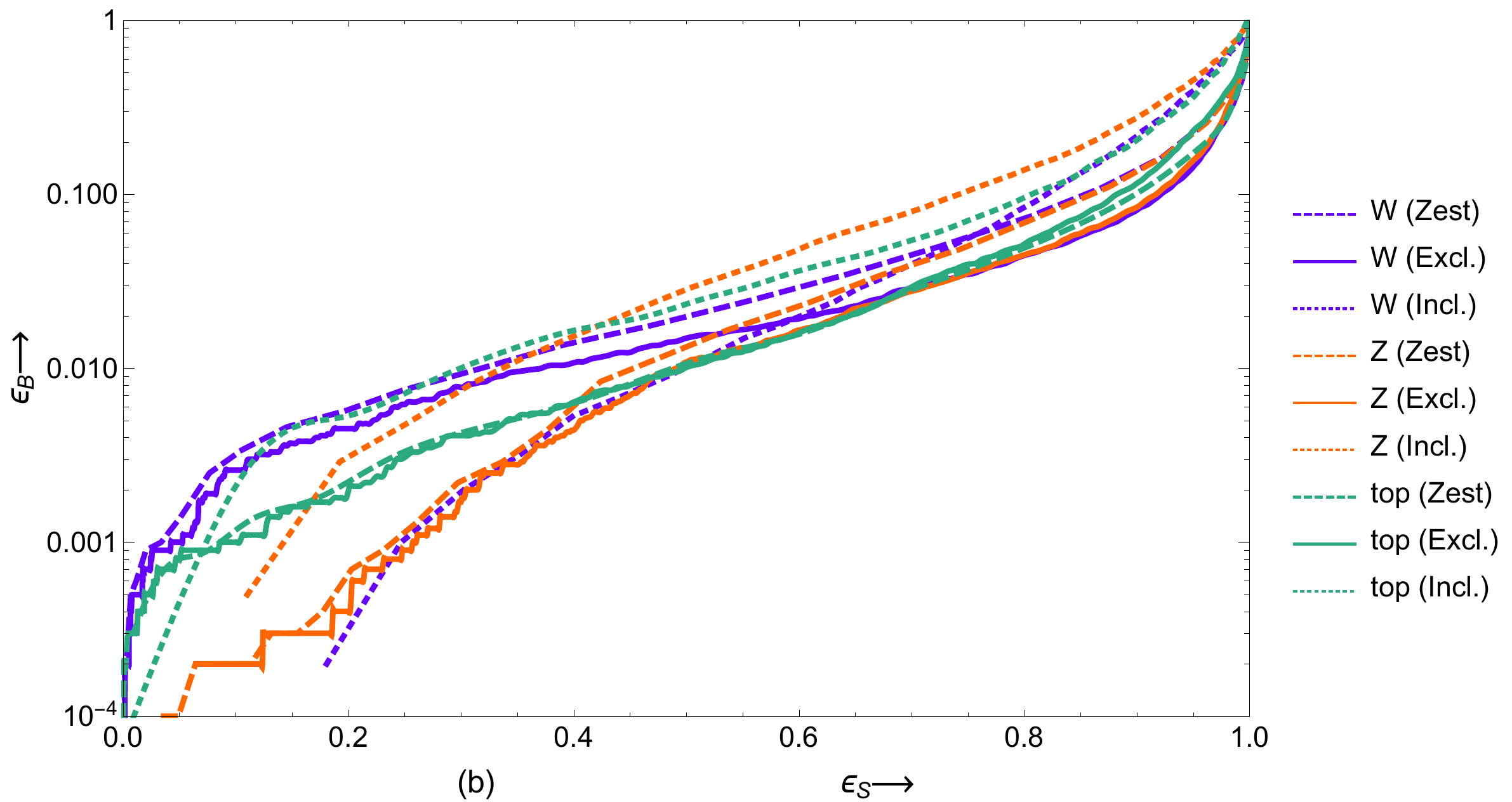} 
\caption{Zest distribution for $W$, $Z$ and top quark along with corresponding gluon jets.}
 \end{figure}
We note that the gluon zest distribution peaks around $\zeta \sim 0.1$ while heavy particle jets are dominantly distributed at higher values of $\zeta$. The near jet-mass independence of gluon zest distribution makes it very useful for vetoing gluon jets. In Fig. 1(b), we have presented the relative operating characteristic curves (dashed lines) for zest based filter, if it alone was used for heavy particle jet identification. We note that, zest provides good signal statistics after high background rejection, for example, nearly 90\% of signal stays in the accepted sample after a zest cut to remove 90\% of gluon jets.
 
 \begin{figure}[t]
 \centering
 \includegraphics[width=0.32\textwidth]{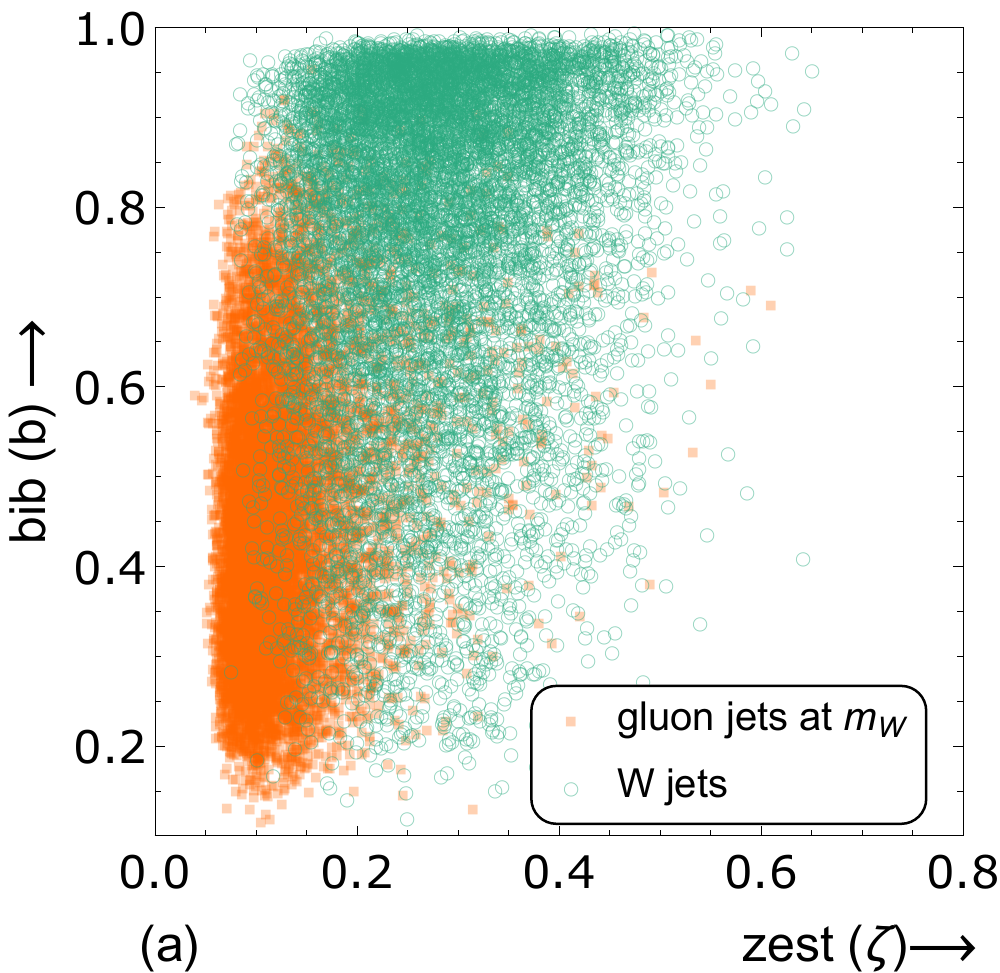}
 \includegraphics[width=0.32\textwidth]{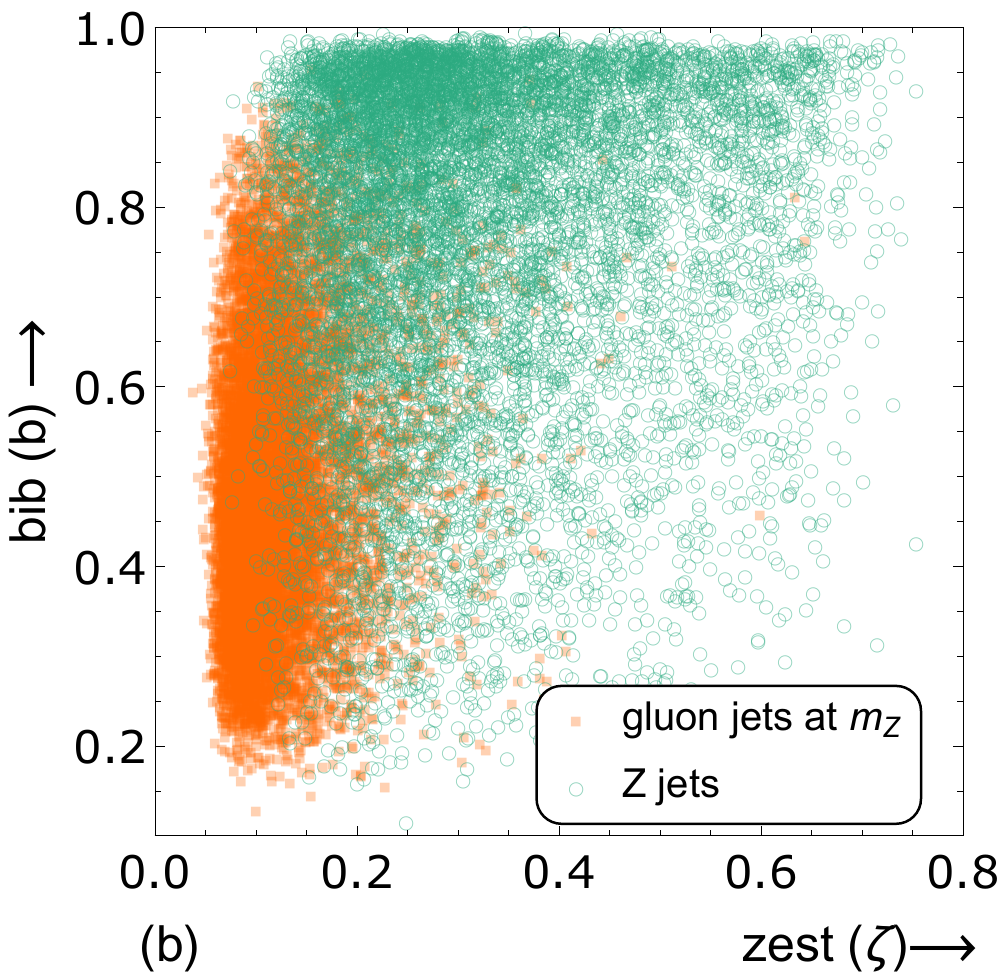}
 \includegraphics[width=0.32\textwidth]{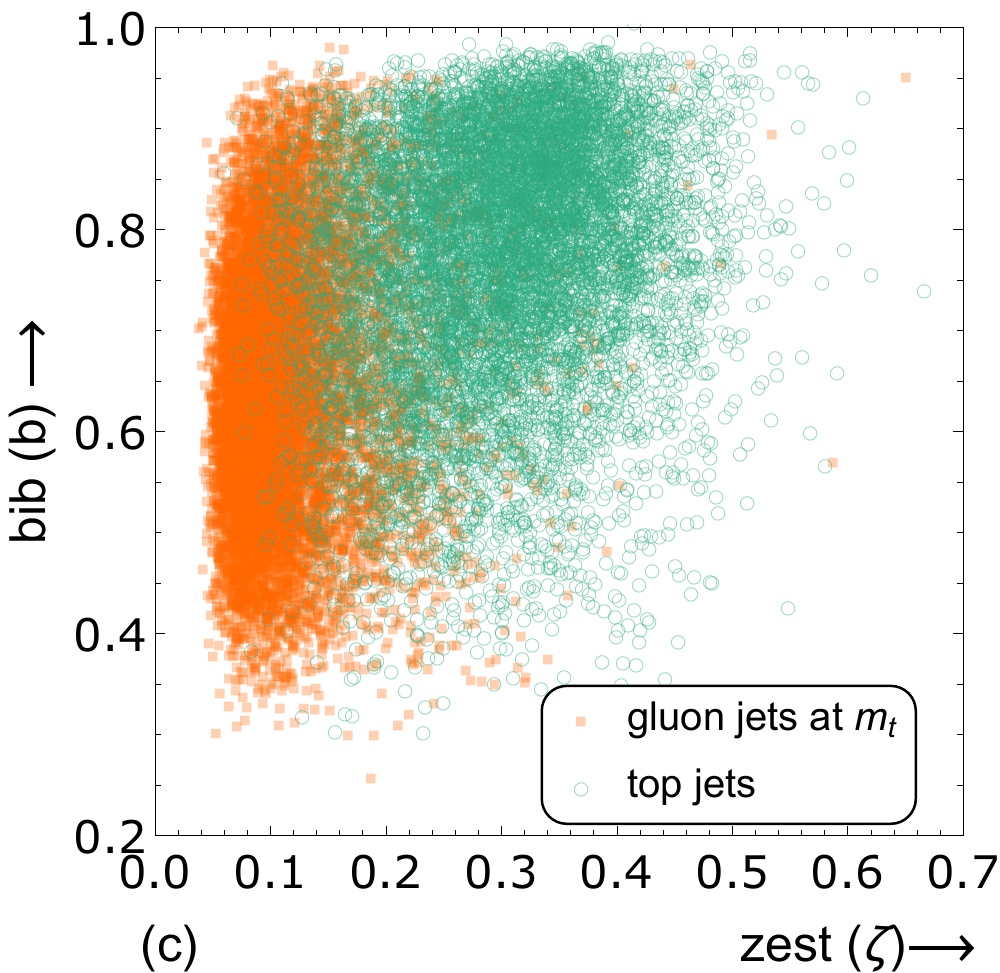}
\caption{(a)-(c) Scatter plots: each point represents a jet}
 \end{figure}
Discrimination for the originating particle can be further improved with multivariate analysis. We introduce a simple infrared safe observable {\em boost-invariant broadening} ({\em bib} for brevity), defined by 
$b = \frac{1}{m_{\rm Jet}}\sum_{i \in {\rm Jet}} \vert \vec p_{\perp i} \vert  \, .$
 Bib is similar to jet broadening defined in \cite{broadening}, however it is invariant under boosts along the jet direction. 
 In Fig.2 (a) to (c), we show scatter plot of simulated jets in zest-bib plane. Gluons occupy small zest and smaller bib region, while heavy particles occupy a large zest and large bib region, thus providing a strong statistical discrimination between gluon jets and heavy particle jets. In Fig. 1 (b), we present three types of relative operating characteristic (ROC) curves (a) zest-based cuts only $-$ represented by dashed lines (b) cuts based on inclusion zones constructed by drawing contours of equal heights on the heavy particle scatter plot $-$ represented by dotted lines, and (c) cuts based on exclusion zones constructed by drawing contours of equal height on the gluon scatter plot $-$ represented by solid lines. We note that for high signal rate, exclusion zone statistics provide slightly better discrimination than only-zest-based cuts for weak bosons, however they provide no significant improvement for top quark. On the other hand, inclusion zone statistics is more efficient when high gluon rejection is required and small signal rate is acceptable.
%

%
%
Although results presented here are preliminary, due to salient properties of zest, we expect that these results will remain mostly unaffected. A complete study will be performed in a future work. \\
\indent \small{AJ thanks SERB, DST for providing support through {\em Ramanujan Fellowship} }.
%
%

%
%


\begin{thebibliography}{6}
%





\bibitem{all-jet-taggers}
Kasieczka et. al., JHEP 1506 (2015) 203. 
Thaler et. al., JHEP 1103 (2011) 015. 
Plehn et. al., Phys.Rev.Lett. 104 (2010) 111801. 
Larkoski et. al., JHEP 1605 (2016) 117. 

\bibitem{jet-grooming}
Ellis et. al., Phys.Rev. D81 (2010) 094023. 
Krohn et. al., JHEP 1002 (2010) 084. 
Dasgupta et. al., Eur.Phys.J. C73 (2013) no.11, 2623. 
Dasgupta et. al., JHEP 1508 (2015) 079. 
Larkoski et. al., JHEP 1405 (2014) 146. 
Butterworth et. al., Phys.Rev.Lett. 100 (2008) 242001. 


\bibitem{zeal-transverse}
R.~Gavai et. al.:
  arXiv:1509.04671 [hep-ph]

\bibitem {future}
Budhraja, A., Jain, A.: under preparation (2017).

\bibitem{pythia}
Sjostrand, T. et. al.: Comput.Phys.Commun. 178 (2008) 852-867. 

\bibitem{thrust}
 E.~Farhi:
   Phys.\ Rev.\ Lett.\  {\bf 39} (1977) 1587.

\bibitem{broadening}
  S.~Catani et. al.:
  Phys.\ Lett.\ B {\bf 295}, 269 (1992).



\end{thebibliography}
\end{document}